\documentclass[aps,pra,twocolumn,superscriptaddress]{revtex4}

\usepackage{graphicx}
\usepackage{amsmath}
\usepackage{amssymb}

\begin{document}

\title{Kondo Effect in Alkali-Earth Atomic Gases with Confinement-induced Resonance }
\author{Ren Zhang}
\affiliation{Institute for Advanced Study, Tsinghua University, Beijing, 100084, China}
\author{Deping Zhang}
\affiliation{Institute for Advanced Study, Tsinghua University, Beijing, 100084, China}
\author{Yanting Cheng}
\affiliation{Institute for Advanced Study, Tsinghua University, Beijing, 100084, China}
\author{Wei Chen}
\affiliation{Institute for Advanced Study, Tsinghua University, Beijing, 100084, China}
\author{Peng Zhang}
\email{pengzhang@ruc.edu.cn}
\affiliation{Department of Physics, Renmin University of China, Beijing, 100872,
China}
\affiliation{Beijing Key Laboratory of Opto-electronic Functional Materials \&
Micro-nano Devices, 100872 (Renmin University of China)}
\author{Hui Zhai}
\email{hzhai@tsinghua.edu.cn}
\affiliation{Institute for Advanced Study, Tsinghua University, Beijing, 100084, China}

\date{\today }

\begin{abstract}
Alkali-earth atoms have a long-lived electronic excited state, and when atoms in this excited state are localized in the Fermi sea of ground state atoms by an external potential, they serve as magnetic impurities, due to the spin-exchange interaction between the excited and the ground state atoms. This can give rise to the Kondo effect. However, in order to achieve this effect in current atomic gas experiment, it requires the Kondo temperature to be increased to a sizable portion of the Fermi temperature. In this paper we calculate the confinement-induced resonance (CIR) for spin-exchanging interaction between the ground and the excited states of the alkali-earth atoms, and we propose that the spin-exchange interaction can be strongly enhanced by utilizing the CIR. We analyze this system by the renormalization group approach, and we show that nearby a CIR, the Kondo temperature can be significantly enhanced. 

\end{abstract}

\maketitle
\section{introduction}
Cold alkali-earth atomic gases have been widely used for building atomic clocks, with which the record of the most accurate optical lattice clock has been achieved \cite{Jun}. This is because alkali-earth atoms have a very long-lived excited ${}^3P_0$ state whose single-particle lifetime can be as long as many seconds. This excited state ${}^3P_0$ and the ground ${}^1S_0$ state are viewed as two internal states of the ``orbital" degree of freedom. Recently, there is an increasing experimental interest in studying many-body physics with alkali-earth atoms, including the $SU(N)$ symmetric interaction and the orbital degree of freedom \cite{Takahashi, Fallani, Jun1, Munich, Florence,Ray}. In particular, recent experiments have demonstrated the inter-orbital spin-exchanging scattering between the ground state ${}^1S_0$ and this ${}^3P_0$ state in fermionic ${}^{88}$Sr \cite{Jun1} and ${}^{173}$Yb atoms \cite{Munich, Florence}.  

\begin{figure}[tp]
\includegraphics[width=2.2 in]
{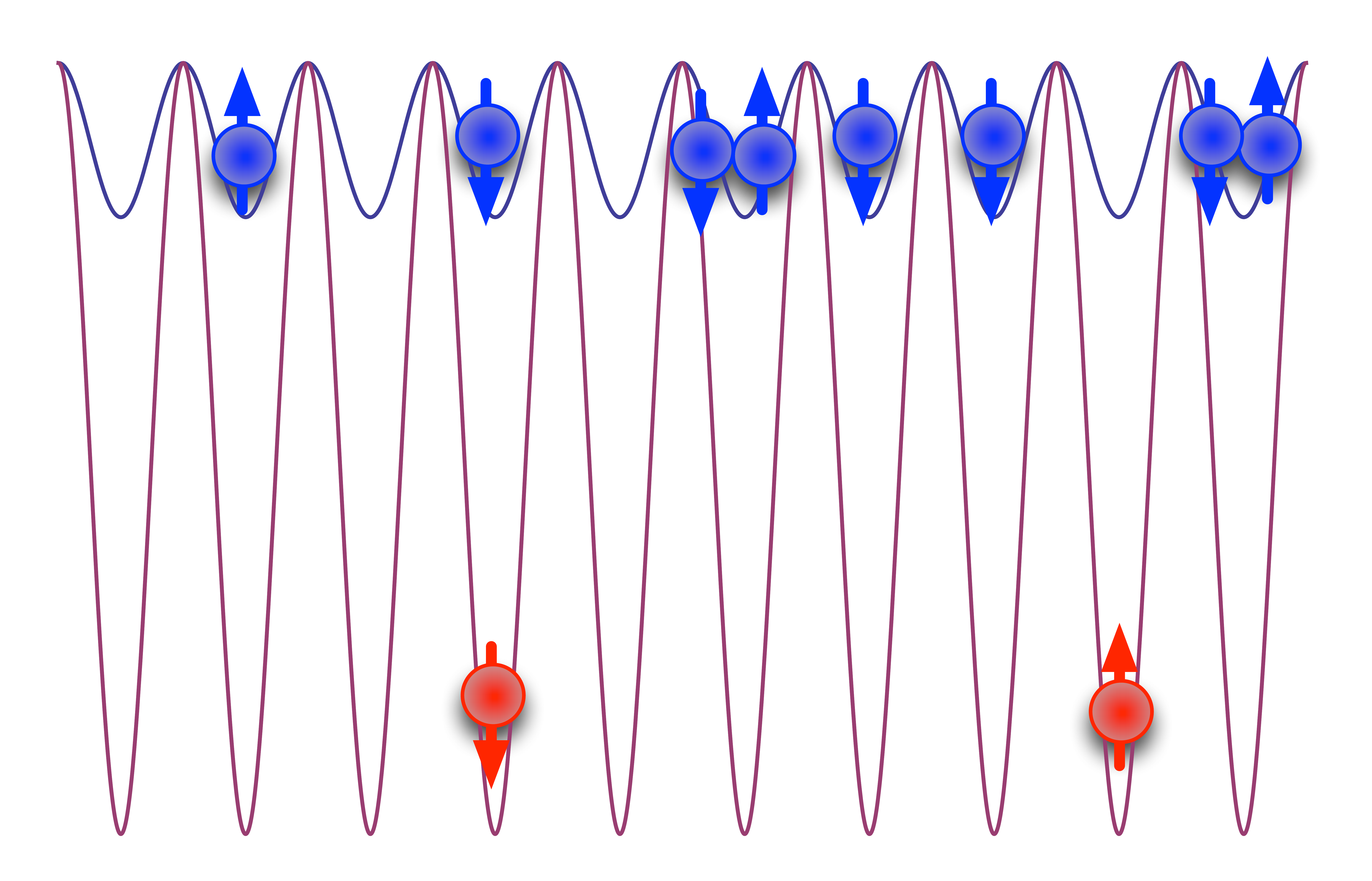}
\caption{Schematic of the system under consideration. The red balls are alkali-earth atoms in ${}^3P_0$ state. They are trapped in a deep lattice and are localized, and have a lower density. The blue balls are alkali-earth atoms in ${}^1S_0$ state. They are trapped in a shallow lattice and are itinerant. The arrows denote the nuclear spin degree of freedom. The system is confined into a one-dimensional tube. }
\label{schematic}
\end{figure}

Utilizing different AC polarizability of ${}^1S_0$ and ${}^3P_0$ states, one can realize the situation that atoms in the ${}^3P_0$ state experience a deep lattice and are localized, while atoms in the ${}^1S_0$ states experience a shallow lattice and remain itinerant, as shown in Fig. \ref{schematic}. Therefore, due to the spin-exchanging scattering between these two states, atoms in the ${}^3P_0$ state can play a role as magnetic impurities in the Fermi sea of atoms in the ${}^1S_0$ state, which can give rise to the Kondo effect \cite{Kondo_alkali_earth}. Realizing the Kondo effect with cold atoms \cite{Kondo_alkali_earth, Kondo1,Kondo2,Kondo3,Kondo4,Kondo5,Yusuke,Demler,Avishai,Rey2} can add a few new ingredients to the Kondo physics, such as the $SU(N)$ Kondo model, manifestation of the Kondo effect other than transport properties and non-equilibrium dynamics. When the magnetic coupling is much weaker comparing to the Fermi energy, such as in the cases of solid-state materials, the Kondo temperature is a few orders of magnitude lower than the Fermi temperature. However, with current cooling power, normally an atomic Fermi gas can only be cooled to $\sim 0.1 T_\text{F}$ ($T_\text{F}$ denotes the Fermi temperature). Therefore, the most challenging question is how to increase the Kondo temperature to the range attainable by current experiments.

In this paper we propose a scheme to overcome this challenge by using confinement-induced resonance (CIR). The CIR phenomenon describes resonant enhancement of the one-dimensional effective interaction strength when a system is confined into a quasi-one-dimensional tube \cite{CIR1,CIR2}. This phenomenon has been observed in previous cold atom experiments with alkali atoms \cite{CIR_ex}.  Here we generalize the CIR phenomenon to the inter-orbital scattering between  ${}^1S_0$ and ${}^3P_0$ states of alkali-earth atoms. We will show that a CIR can strongly enhance the spin-exchanging scattering, and consequently, the Kondo temperature can be increased to a sizable fractional of the Fermi temperature when a CIR is approached. 

\section{Confinement Induced Resonance}

\subsection{Zero-Magnetic Field} To illustrate the basic ideas, we first discuss a two-body problem at zero magnetic field. Let us briefly review interaction between two fermions in two different orbital states ${}^3P_0$ (denoted by $|e\rangle$) and ${}^1S_0$ (denoted by $|g\rangle$) and different nuclear spin states (for simplicity, here we only take two nuclear spin states   denoted by $|\uparrow\rangle$ and $|\downarrow\rangle$), respectively. We can introduce four antisymmetric bases for the internal degrees of freedom 
\begin{align}
&|\pm\rangle=\frac{1}{2}(|ge\rangle\pm |eg\rangle)(|\uparrow\downarrow\rangle\mp|\downarrow\uparrow\rangle)\\
&|g\uparrow;e\uparrow\rangle=\frac{1}{\sqrt{2}}(|g e \rangle-|e  g\rangle)|\uparrow\uparrow\rangle\\
&|g\downarrow;e\downarrow\rangle=\frac{1}{\sqrt{2}}(|g e \rangle-|e  g\rangle)|\downarrow\downarrow\rangle)
\end{align} 
in which $s$-wave scattering is allowed. $|+\rangle$ is orbital triplet and nuclear spin singlet, and the other three are orbital singlet and nuclear spin triplet. Since nuclear spin does not participate the inter-atomic interaction process, the interaction part possesses the nuclear spin rotational symmetry, for which spin singlet and triplet will not mix and the interaction potentials are the same for $|-\rangle$ and $|g\uparrow;e\uparrow\rangle$ and $|g\downarrow;e\downarrow\rangle$ channels. Therefore, the inter-atomic potential $\hat{V}({\bf r})$ is diagonal in the bases $\{|+\rangle,|-\rangle,|g\uparrow;e\uparrow\rangle,|g\downarrow;e\downarrow\rangle\}$ as
\begin{equation}
V_+({\bf r})\mathcal{P}_++V_-({\bf r})(\mathcal{P}_{-}+\mathcal{P}_{\uparrow\uparrow}+\mathcal{P}_{\downarrow\downarrow}),\nonumber
\end{equation}
where $\mathcal{P}_i=|i\rangle\langle i|$, $i=\pm$, and $\mathcal{P}_{\uparrow\uparrow}=|g\uparrow;e\uparrow\rangle\langle g\uparrow;e\uparrow|$ and $\mathcal{P}_{\downarrow\downarrow}=|g\downarrow;e\downarrow\rangle\langle g\downarrow;e\downarrow|$. The two interaction potentials are denoted by $V_\pm({\bf r})=2\pi \hbar^2 a_\pm\delta({\bf r})\frac{\partial}{\partial r}(r\cdot)/\mu$ ($\mu$ is the two-body reduced mass), and $a_\pm$ are two independent scattering lengths. 

We can rotate the interaction potential $\hat{V}({\bf r})$ into another bases $\{|g\uparrow;e\downarrow\rangle, |g\downarrow;e\uparrow\rangle, |g\uparrow;e\uparrow\rangle, |g\downarrow;e\downarrow\rangle\}$, where $|g\uparrow;e\downarrow\rangle=(1/\sqrt{2})(|+\rangle+|-\rangle)$ and $|g\downarrow;e\uparrow\rangle=(1/\sqrt{2})(|-\rangle-|+\rangle)$, and $V({\bf r})$ becomes 
\begin{align}
\hat{V}&=\frac{V_++V_-}{2}(\mathcal{P}_{\uparrow\downarrow}+\mathcal{P}_{\downarrow\uparrow})+\frac{V_--V_+}{2}(\mathcal{S}_\text{ex}+\mathcal{S}^\dag_\text{ex})\nonumber \\
&+V_{-}(\mathcal{P}_{\uparrow\uparrow}+\mathcal{P}_{\downarrow\downarrow}),
\end{align}
where $\mathcal{P}_{\uparrow\downarrow}=|g\uparrow;e\downarrow\rangle\langle g\uparrow;e\downarrow|$, $\mathcal{P}_{\downarrow\uparrow}=|g\downarrow;e\uparrow\rangle\langle g\downarrow;e\uparrow|$, and $\mathcal{S}_\text{ex}=|g\uparrow;e\downarrow\rangle\langle g\downarrow;e\uparrow|$. In the presence of a lattice as described in Fig. \ref{schematic}, when atoms in the $|e\rangle$ state are localized as impurities while atoms in the $|g\rangle$ state remain itinerant, the off-diagonal $(V_--V_+)/2$ represents the process that an itinerant fermion exchanges its spins with impurities, and this spin-exchanging process is the essential process responsible for the Kondo effect \cite{Kondo}. Normally, this interaction strength is much smaller comparing to the Fermi energy, and the Kondo temperature is exponentially suppressed \cite{Kondo}.

Now let us consider atoms confined into a quasi-one-dimensional tube by a transverse harmonic trap. Here we consider the situation that the transverse confinement is the same for both ${}^3P_0$ and ${}^1S_0$ states, which can be achieved by applying a two-dimensional optical lattice in the $xy$-plane with the magic wave length \cite{magic_wave}. And we first consider the situation without lattice along the longitudinal $z$-direction. The free-Hamiltonian $\hat{H}_0$ can be separated into the center-of-mass part and the relative motion part $\hat{H}_\text{r}$, where
\begin{equation}
\hat{H}_\text{r}=-\frac{\hbar^2}{2\mu}\nabla^2+\frac{\mu\omega^2}{2}(x^2+y^2).
\end{equation} 
For single interaction channel with a three-dimensional scattering length $a_\text{s}$, it is known that the interaction strength of the effective one-dimensional interaction potential $g_{0}\delta(z)$ is given by \cite{CIR1}
\begin{equation}
g_{0}=\frac{2\pi \hbar^2 a_\text{s}}{\mu}|\phi_{00}|^2\left(1-\mathcal{C}\frac{a_\text{s}}{a_\perp}\right)^{-1}, \label{CIR}
\end{equation}
where $\phi_{00}$ is the ground state wave function of $\hat{H}_\text{r}$, $a_\perp=\sqrt{\mu\omega/\hbar}$ is the harmonic length and $\mathcal{C}=1.4603\cdots$ is a constant. $g$ diverges when $a_\perp=\mathcal{C}a_\text{s}$, which is known as CIR \cite{CIR1}. This resonance occurs when the energy of a bound state in the transverse excited modes matches the scattering threshold \cite{CIR2}.   

\begin{figure}[tp]
\includegraphics[width=3. in]
{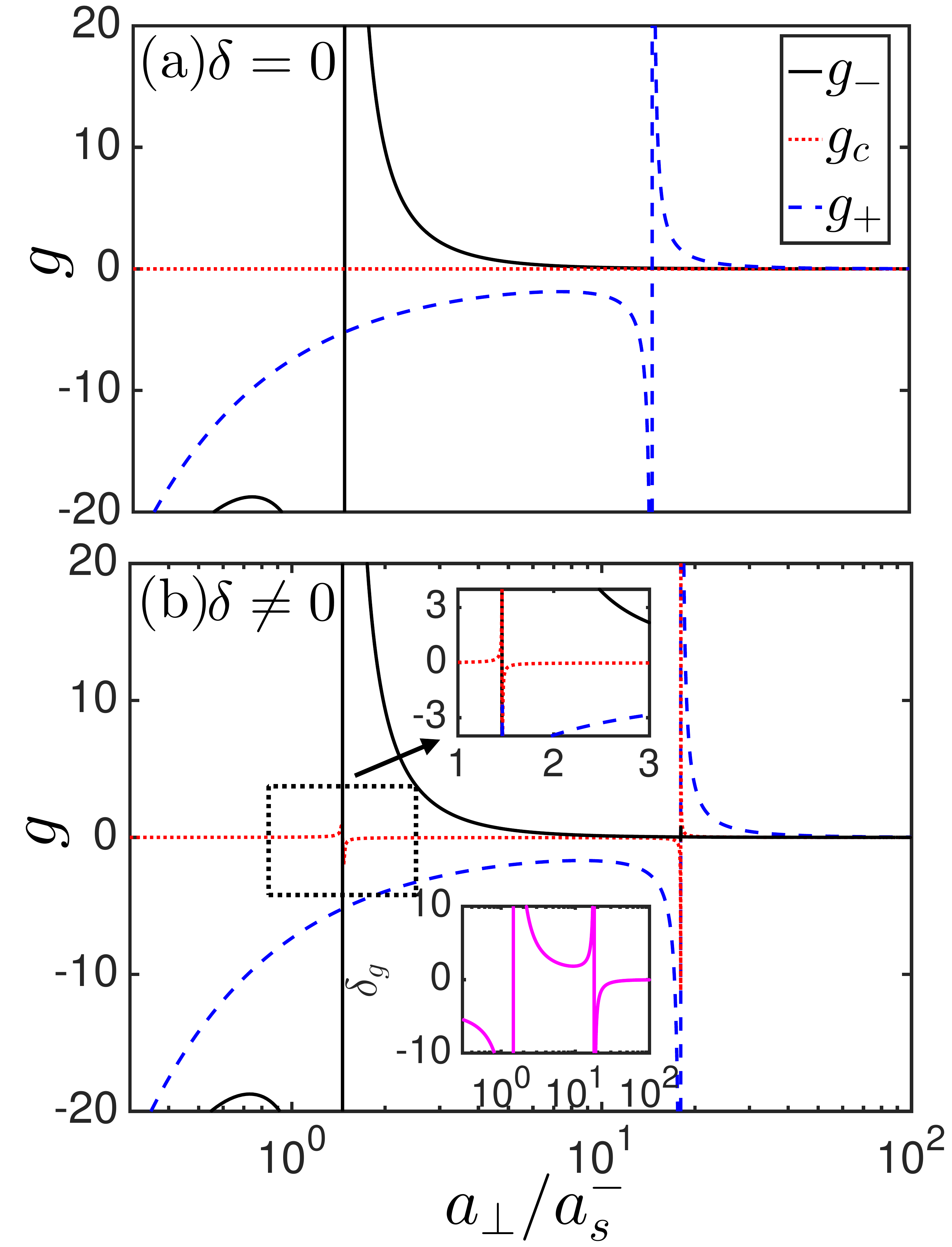}
\caption{The one-dimensional interaction strength $g_{+}$, $g_{-}$ and $g_\text{c}$($\hbar^{2}/(\mu10^{3}a_{0})$ is taken as unit) as a function of $a_\perp/a_{s}^{-}$. (a) The magnetic field $B=0$. Here $g_{c}=0$. (b) $B=35G$. Here we use ${}^{173}$Yb as an example and take $\delta=2\pi\hbar 112\times B$, and therefore $\delta=25\hbar k$Hz at $35G$. In both cases, $g_{0}$ always behaves the same as $g_{-}$ in case (a). For ${}^{173}$Yb we take $a_{+}\simeq10 a_{-}$ and $a_{-}=200a_0$ \cite{OFR-exp1,OFR-exp2}. The first CIR takes place around $a_\perp\approx\mathcal{C}a_{-}$ and the second CIR takes place around $a_\perp\approx\mathcal{C}a_{+}$, though the exact locations of CIRs will be shifted by magnetic field. The insets in (b) show $g$'s nearby the first CIR and $\delta g=g_{-}-g_{+}$. }
\label{CIR_F}
\end{figure}

At zero-field, it is easy to show that $|+\rangle$, $|-\rangle$, $|g\uparrow;e\uparrow\rangle$ and $|g\downarrow;e\downarrow\rangle$ are all eigenstates of free Hamiltonian $\hat{H}_0$. Hence, under confinement, these four scattering channels can be treated as independent channels, and the reduced one-dimensional interaction still takes a diagonal form as
\begin{align}
\hat{V}_\text{1D}=[g_+\mathcal{P}_{+}+g_-\mathcal{P}_{-}+g_{0}\mathcal{P}_{\uparrow\uparrow}+g_{0}\mathcal{P}_{\downarrow\downarrow}]  \delta(z), \label{zeroB}
\end{align}
where $g_{+}$ is related to $a_{+}$, and $g_{-}$, $g_{0}$ are related to $a_{-}$ via Eq. (\ref{CIR}), respectively. Therefore, when rotated to $\{|g\uparrow;e\downarrow\rangle, |g\downarrow;e\uparrow\rangle, |g\uparrow;e\uparrow\rangle, |g\downarrow;e\downarrow\rangle\}$ bases, the spin-exchange interaction term is now given by $(g_--g_+)\delta(z)/2$. When $a_\perp\rightarrow \mathcal{C} a_{+}$, $g_+$ diverges and $g_{-}$ remains finite; and when $a_\perp\rightarrow \mathcal{C} a_{-}$, $g_{-}$ diverges and $g_{+}$ remains finite, as shown in Fig. \ref{CIR_F}(a). Therefore, the spin-exchanging interaction becomes very strong nearby these two CIRs, and consequently, the Kondo temperature can be dramatically enhanced. This is the basic idea of our proposal.

\subsection{Finite Magnetic Field} In real experiment, there is always a finite magnetic field. Due to the difference in the Land\'e g-factor between $|g\rangle$ and $|e\rangle$ states, $|g\uparrow;e\downarrow\rangle$ and $|g\downarrow;e\uparrow\rangle$ states differ by a finite energy $\delta$ proportional to the magnetic field strength \cite{Jun_PRA}. In another word, this leads to a mixing term between $|+\rangle$ and $|-\rangle$ given by $\delta/2(|+\rangle\langle -|+h.c.)$ \cite{Munich, Florence}. Therefore, these two channels can no longer be treated independently, which brings out additional complication to our proposal. Under the condition $\delta\ll \hbar\omega$, both two channels are retained in the one-dimensional effective mode. To deduce the one-dimensional interactional strength, our calculation follows the standard procedure of CIR discussed before \cite{CIR1}, that is, one first obtains an effective one-dimensional scattering amplitude in both $|g\uparrow;e\downarrow\rangle$ and $| g\downarrow;e\uparrow \rangle$ channels by solving three-dimensional Hamiltonian with the confinement potential, and then constructs an effective one-dimensional model which gives exactly the same scattering amplitude.

The Hamiltonian for the relative motion of the two-body system with a confinement potential can be written as 
\begin{align}
\hat{H}_{c}&=\left[-\frac{\hbar^{2}}{2\mu}\nabla_{{\bf r}}^{2}+\frac{\mu\omega^{2}}{2}\left(x^{2}+y^{2}\right)\right]({\cal P}_{+}+{\cal P}_{-})+\frac{\delta}{2}({\cal S}_{c}+{\cal S}_{c}^{\dagger})\nonumber\\
&+V_{+}({\bf r}){\cal P}_{+}+V_{-}({\bf r})({\cal P}_{-}+\mathcal{P}_{\uparrow\uparrow}+\mathcal{P}_{\downarrow\downarrow}),\label{eqn:rel-H}
\end{align}
where ${\cal S}_{c}=|+\rangle\langle-|$. Two important features at finite $\delta$ are worth emphasizing here. First, in the three-dimension, interaction does not mix $|+\rangle$ and $|-\rangle$ because it respects the nuclear spin rotational symmetry. However, the finite Zeeman-field $\delta$ in the single-particle Hamiltonian breaks the nuclear spin rotational symmetry. In the quasi-one-dimensional system, since the virtual processes to the transverse excited levels are taken into account when reducing dimensionality, the effect of this Zeeman energy term enters the effective one-dimensional interaction term $\hat{V}_{\text{1D}}$ through the intermediate state energy of these virtual processes. Hence, the effective 1D Hamiltonian is expected to take the form as
\begin{eqnarray}
\hat{H}_{\rm 1D}=\left(-\frac{\hbar^{2}}{2\mu}\frac{d^{2}}{dz^{2}}\right)({\cal P}_{+}+{\cal P}_{-})+\frac{\delta}{2}({\cal S}_{c}+{\cal S}_{c}^{\dagger})+V_{1D},\label{eqn:eff-H}
\end{eqnarray}
where
\begin{align}
\hat{V}_{\text{1D}}=[g_+\mathcal{P}_{+}+g_{-}\mathcal{P}_{-}+g_\text{c}(\mathcal{S}_\text{c}+\mathcal{S}^\dag_\text{c})
+g_{0}(\mathcal{P}_{\uparrow\uparrow}+\mathcal{P}_{\downarrow\downarrow})]\delta(z). \label{finiteB}
\end{align}
It should be noticed that $V_{1D}$ no longer respects the nuclear spin rotational symmetry, and contains a non-zero off-diagonal term $g_\text{c}$  that couples $|+\rangle$ and $|-\rangle$ channels. Similar effect has also been discussed in the CIR of spinor atoms \cite{Cui}. $g_{+}$, $g_{-}$ and $g_\text{c}$ depend on the strength of the Zeeman field, as shown in Fig. \ref{CIR_F}(b). The detailed derivation of these parameters are given in the appendix. While  $g_{0}$ are still related to $a_{-}$ via Eq. (\ref{CIR}) since $|g\uparrow;e\uparrow\rangle$ and $|g\downarrow;e\downarrow\rangle$ states are still eigenstates despite of the presence of the Zeeman term. 

Secondly, when $g_{+}$ diverges, $g_{-}$ and $g_\text{c}$ will also diverge, as shown in Fig. \ref{CIR_F}(b). Similar happens when $g_{-}$ diverges. After the base rotation, the interaction term becomes
\begin{align}
&\hat{H}_{\text{int}}=\left[\left(\frac{g_{+}+g_{-}}{2}+g_\text{c}\right) \mathcal{P}_{\uparrow\downarrow}+\left(\frac{g_{+}+g_{-}}{2}-g_\text{c}\right) \mathcal{P}_{\downarrow\uparrow}\right. \nonumber\\
&\left.+\frac{g_{-}-g_{+}}{2}\left(\mathcal{S}_\text{ex}+\mathcal{S}^\dag_\text{ex}\right)+g_{0}\left(\mathcal{P}_{\uparrow\uparrow}+\mathcal{P}_{\downarrow\downarrow}\right) \right]\delta(z), \label{finiteB2}
\end{align} 
where the spin-exchanging term is still given by $(g_--g_+)\delta(z)/2$. Fortunately, as shown in Fig. \ref{CIR_F}(b), one can see that, nearby one of the CIRs, either $g_{-}$ diverges much slower than $g_{+}$, or $g_{+}$ diverges much slower than $g_{-}$. Therefore $(g_--g_+)/2$ still displays a divergent behavior and the insight gained from the zero-field limit will stay hold.

\section{The Kondo Effect}
\subsection{Lattice Model with a Single Impurity} Now we consider turning on the lattice potential as shown in Fig. \ref{schematic}, which localizes atoms in the $|e\rangle$ state as impurities in a Fermi sea of atoms in the $|g\rangle$ state. We also consider the regime where the density of impurity atoms is much more dilute than the density of itinerant atoms, and for simplicity, we consider a single impurity problem. The tight-binding model is given by
\begin{align}
&\hat{H}=\hat{H}_0+\hat{H}_\text{I}\\
&\hat{H}_0=\sum\limits_{k,\sigma}(-t\cos k)c^\dag_{k\sigma}c_{k\sigma}+\delta S^z/2-\frac{\delta}{2L}\sum\limits_{k}s^z_{kk}\\ 
&\hat{H}_{\rm I}=\frac{1}{L}\sum_{k,q}\left\{\frac{J_{+}}{2}S^{+}s^-_{kq}+\frac{J_{-}}{2}S^{-}s^+_{kq}+\frac{J_{z}}{2}S^{z}s^z_{kq}\right.\nonumber\\
&\left.+ Un_{kq}+\frac{U_1}{2}S^{z}n_{kq}
+U_{2}s^z_{kq}\right\}, \label{lattice_model}
\end{align}
where $c_{k\sigma}$ and $d_\sigma$ are fermion operators for itinerant fermions and impurity fermion, respectively, $S^{+}=d_{\uparrow}^{\dagger}d_{\downarrow}$,  $S^{-}=d_{\downarrow}^{\dagger}d_{\uparrow}$, $S^{z}=(1/2)(d_{\uparrow}^{\dagger}d_{\uparrow}-d_{\downarrow}^{\dagger}d_{\downarrow})$, $s^-_{kq}=c_{k,\downarrow}^{\dagger}c_{q,\uparrow}$, $s^+_{kq}=c_{k,\uparrow}^{\dagger}c_{q,\downarrow}$, $s^z_{kq}=(1/2)(c_{k,\uparrow}^{\dagger}c_{q,\uparrow}-c_{k,\downarrow}^{\dagger}c_{q,\downarrow})$ and $n_{kq}=c_{k,\uparrow}^{\dagger}c_{q,\uparrow}+c_{k,\downarrow}^{\dagger}c_{q,\downarrow}$. $t$ is the hopping amplitude of itinerant fermions. The interaction parameters $J_\perp$, $J_z$, $U_z$, $U_{+}$ and $U_{-}$ are related to $g_{+}$, $g_{-}$ and $g_\text{c}$ via
\begin{align}
&J_{\pm}\propto-(g_{+}-g_{-}),\hspace{0.2cm}J_{z}\propto-(g_{+}+g_{-}-2g_0), \label{Jcoeff}\\
&U \propto \frac{g_{+}+g_{-}+2g_0}{4}, \hspace{0.2cm} U_{1}\propto -2g_\text{c}, \hspace{0.2cm}U_{2}\propto g_\text{c},
\end{align}
where the proportional constant depends on the details of Wannier function overlap between localized atoms and itinerant atoms (In the plot of following figures, this constant has been chosen as $0.03$).

Here we first discuss a few features of this lattice model: 

1. The first three terms ($J_\pm$-and $J_z$-terms) in $\hat{H}_\text{I}$ describe the Kondo coupling. From Fig. \ref{CIR_F} one can see that i) in the left side of the first CIR, and the right side of the second CIR,  $J_\pm, J_z<0$ and ii) in the right side of the first CIR, and in the left side of the second CIR, $J_\pm, J_z>0$. Thus, both ferromagnetic and anti-ferromagnetic Kondo couplings can be accessed by tuning the confinement length $a_\perp$. At zero-field, when $g_0=g_{-}$, from Eq. (\ref{Jcoeff}) we have $J_{\pm}=J_z$ and the Kondo coupling respects spin-rotational symmetry, while at finite Zeeman field, generally $g_0$ is not equal to $g_{-}$ and it gives rise to an anisotropic Kondo model. 

2. The fourth term ($U$-term) in $\hat{H}_\text{I}$ describes a potential scattering that does not depend on spins. This comes from the diagonal term in Eq. (\ref{finiteB2}). Furthermore, at finite field, due to the absence of the full spin rotational symmetry (a rotational symmetry along spin z-axes is still present), there exist other two scattering terms ($U_1$-and $U_2$-terms) in $\hat{H}_\text{I}$.  It naturally raises the question that whether these extra terms will affect the Kondo physics. 

3. The $\delta$-term in $\hat{H}_0$ comes from the different $g$-factor between localized and itinerant fermions. When this term is sufficiently large, it tends to polarize fermions and will destroy the Kondo physics. 

We shall also remark that here the interaction between itinerant fermions is ignored. Since microscopically it is described by another independent scattering length between atoms in $|g\rangle$ states with different nuclear spins (normally denoted by $a_\text{gg}$.) The CIR for $a_\text{gg}$ is reached at a different confinement radius when $a_\perp=\mathcal{C}a_\text{gg}$, where the interaction between itinerant fermions will become very strong. With interactions, the effect of a magnetic impurity in a Luttinger liquid has been studied before \cite{Lee,Nagosa}, and in the strongly interaction limit the results will be reported elsewhere \cite{Zhai}. Whereas, at these two CIRs we focus on here, this interaction is rather weak and can be safely ignore. 

\begin{figure}[tp]
\includegraphics[width=3.in]{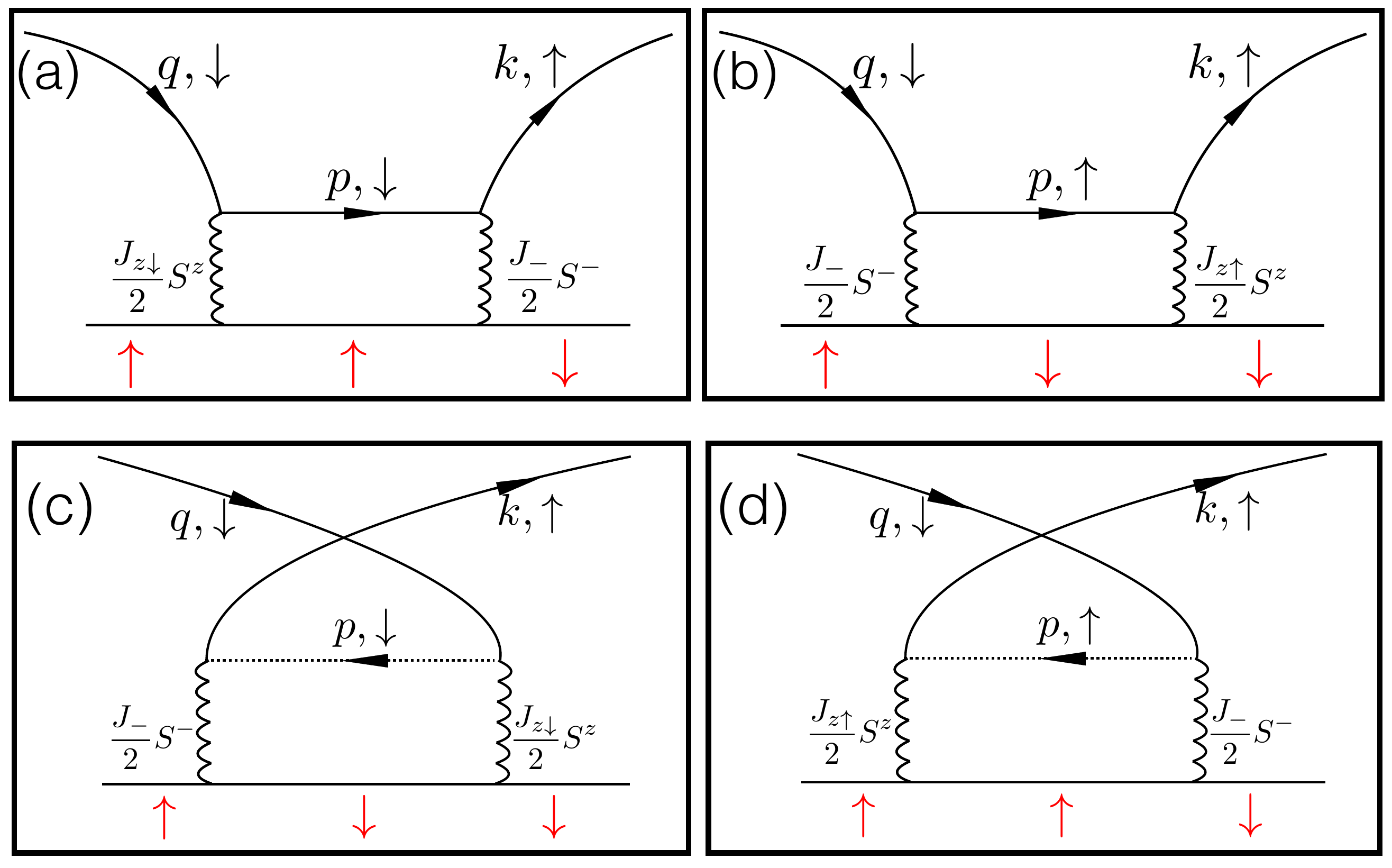} \caption{Second order diagram that can renormalize $J_{-}$. Black arrows denote
the itinerant atomic spin and red arrows denote the impurity atomic
spin. (a) and (b) are ``particle'' process. (c) and (d) are ``hole''
process.}
\label{fig:feymanfig} 
\end{figure}

\subsection{Renormalization Group Studies.} To address the effect of these extra terms and to make a more concrete predication of the Kondo temperature, here we adopt the renormalization group (RG) approach well established for the Kondo problem \cite{Kondo,Anderson}. The key idea is to iteratively integrate out the high-energy modes of itinerant fermions and see how the interaction parameters flow. In this analysis,
we only consider the lowest order virtual processes. For simplicity,
only the scattering processes that renormalize $J_{-}$ are shown
explicitly in Fig.~\ref{fig:feymanfig} as an example. The renormalization to the $J_{-}$ term from the processes in Fig.~\ref{fig:feymanfig}(a)
when the cutoff is reduced from $D$ to $D-\delta D$ reads 
\begin{eqnarray}
 &  & \sum_{p}\frac{J_{-}}{2}S^{-}c_{k,\uparrow}^{\dagger}c_{p,\downarrow}\frac{1}{\omega+\epsilon_{q}-\epsilon_{p}}\frac{J_{z\downarrow}}{2}\ S^{z}c_{p,\downarrow}^{\dagger}c_{q,\downarrow}\nonumber \\
 &  & \approx\frac{J_{z\downarrow}J_{-}}{4}S^{-}S^{z}\rho_{0}|\delta D|c_{k\uparrow}^{\dagger}c_{q\downarrow}\frac{1}{-D}\nonumber \\
 &  & =-\frac{1}{8}J_{z\downarrow}J_{-}\rho_{0}|\delta D|S^{-}c_{k\uparrow}^{\dagger}c_{q\downarrow}\frac{1}{D},\label{eqn:poorman-exa}
\end{eqnarray}
where $\rho_{0}$ is the density of states of the itinerant atoms
near Fermi surface and we have set $c_{p}c_{p}^{\dagger}=1$ for the
$p$ states in the energy scale between $D$ and $D-\delta D$ and
$S^{z}=1/2$ for up-spin impurity. The approximation in Eq.~(\ref{eqn:poorman-exa})
comes from the fact that $\epsilon_{p}\approx D$ and $D\gg\omega,\epsilon_{q}$.
Similarly, one can write the renormalization contribution to the $J_{-}$
term from Fig.~\ref{fig:feymanfig}(b),(c),(d) as 
\begin{align}
&\frac{1}{8}J_{z\uparrow}J_{-}\rho_{0}|\delta D|S^{-}c_{k\uparrow}^{\dagger}c_{q\downarrow}\frac{1}{D+\delta};\\&-\frac{1}{8}J_{z\downarrow}J_{-}\rho_{0}|\delta D|S^{-}c_{k\uparrow}^{\dagger}c_{q\downarrow}\frac{1}{D+\delta};\\&\frac{1}{8}J_{z\uparrow}J_{-}\rho_{0}|\delta D|S^{-}c_{k\uparrow}^{\dagger}c_{q\downarrow}\frac{1}{D},
\end{align}
It turns out that the contribution of the diagrams which
involve the $U_{\sigma}$ term is zero. Thus one can sum up all the
diagrams and obtain the renormalization equation for $J_{-}$ as 
\begin{eqnarray}
\frac{dJ_{-}}{dD}=\frac{\rho_{0}}{4}J_{-}(J_{z\downarrow}-J_{z\uparrow})\left(\frac{1}{D+\delta}+\frac{1}{D}\right).
\end{eqnarray}
By repeating similar procedures, one can find the renormalization
equations of all the interaction parameters in Eq.~(\ref{lattice_model})
as follows: 
\begin{align}
&\frac{dJ_{+}}{dD}=-\frac{\rho_{0}}{2}J_{+}J_{z}\left(\frac{1}{D-\delta}+\frac{1}{D}\right),\\
&\frac{dJ_{-}}{dD}=-\frac{\rho_{0}}{2}J_{-}J_{z}\left(\frac{1}{D+\delta}+\frac{1}{D}\right),\\
&\frac{dJ_{z}}{dD}=-\frac{\rho_{0}}{2}J_{+}J_{-}\left(\frac{1}{D-\delta}+\frac{1}{D+\delta}\right),\\
&\frac{dU_2}{dD}=\frac{\rho_{0}}{4}J_{+}J_{-}\left(\frac{1}{D-\delta}-\frac{1}{D+\delta}\right),\\
&\frac{dU_{1}}{dD}=0,\hspace{0.3cm}\frac{dU}{dD}=0, \label{RG-equation}
\end{align}  
where we have taken $D$ as the energy cutoff, and its initial value is chosen to be an energy scale of the order of the Fermi energy.  And for simplicity, we have also assumed a constant density of state denoted by $\rho_{0}$. When $\delta=0$, Eq.~(\ref{RG-equation}) can reduce to the RG equations derived in Ref. \cite{Anderson}. At this level of approximation, $U$ and $U_1$ terms are renormalized. 

\begin{figure}[tp]
\includegraphics[width=2.5 in]{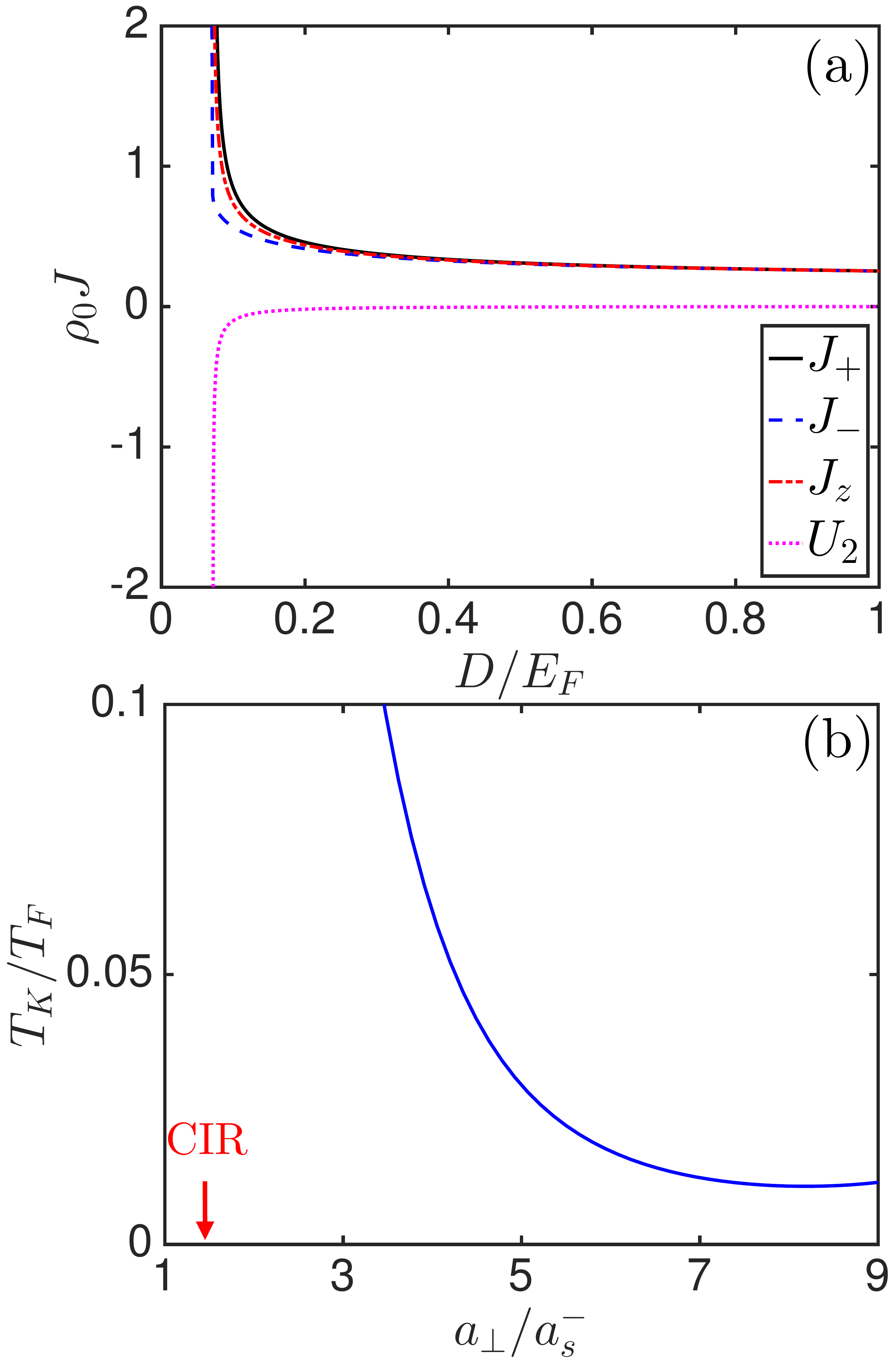}
\caption{(a) Flow of parameters $J_{\pm}$, $J_z$ and $U_2$ as lowering energy cutoff $D$. Here we have taken a typical density that gives rise to $E_\text{F}=5\hbar\times10^4$, and  the initial values of $J_\pm$, $J_z$ and $U_2$ as $\rho_0J_\pm=0.2377$, $\rho_0J_z=0.2376$ and $\rho_0 U_z=-3\times 10^{-4}$. (b) The Kondo temperature increases as the confinement length $a_\perp$ is tuned toward the first CIR from the anti-ferromagnetic coupling side. Here we have taken $\delta=0.07 E_\text{F}$.  }
\label{RG}
\end{figure}

Here we focus on the antiferromagnetic case. In Fig. \ref{RG}(a), we show a typical flow of the RG equations with $\delta\ll E_\text{F}$. We find that as lowering the energy cutoff, $J_\pm$ and $J_z$ term diverge much faster than $U_2$ term. This in fact can already be seen in the RG equation, since $dU_2/dD$ scales with $\delta/D^2$ and therefore evolves very slowly when $\delta\ll D$. Thus, at the energy scale when $J_\pm$, $J_z$ diverge, the Kondo effect appears while the strength of other terms remain quite small. Hence, we conclude that the extra interaction terms in the lattice model Eq. (\ref{lattice_model}) will not affect the Kondo effect in this system. 

The divergent energy scale for $J_\pm$, $J_z$ is normally taken as the Kondo temperature \cite{Kondo}. In Fig. \ref{RG}(b) we show the dramatic increase of the Kondo temperature as approaching one of the CIRs. The Kondo temperature can increase to $\sim 0.1 T_\text{F}$ which is attainable by current experiments. The underlying physics is basically the increasing of spin-exchanging coupling as we discussed above. In the plot of Fig. \ref{RG}(b), we have restricted out initial interaction parameter $\rho_0 J_{\pm}, \rho_0 J_z \lesssim 0.2$.  When it is very close to the CIR, the initial value of these interaction parameters are already very large which invalidates the perturbative RG approach. The impurity physics with a large or even divergent spin-exchanging interaction remains a challenging issue, and this makes the experimental quantum simulation studies of this model even more interesting.  

\begin{figure}[t]
\includegraphics[width=2.7in]{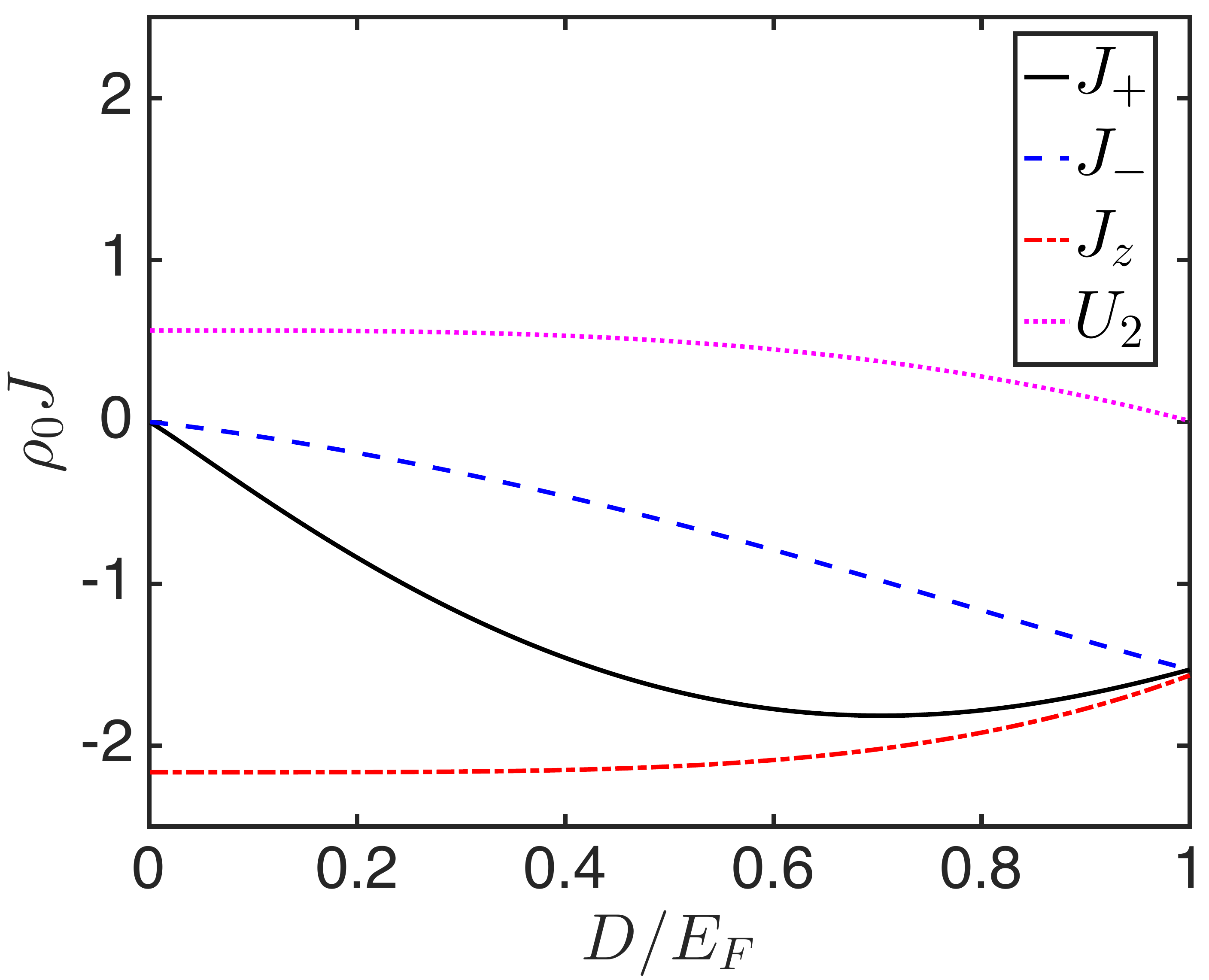} \caption{RG flow of parameters $J_{\pm},J_{z}$ and $U$ in a strong magnetic
field with B=100G and $\delta/E_{F}\simeq1.4$ .}
\label{fig:flow} 
\end{figure}

Solving the RG equations for $\delta\sim E_\text{F}$ or $\delta\gg E_\text{F}$, we find none of the interaction parameters will diverge even when the cutoff $D$ is lowered to zero as shown in Fig.~\ref{fig:flow}, which means the absence of the Kondo effect. Therefore, for a fixed $a_\perp/a_\text{s}$, increasing $\delta$ simply by increasing the magnetic field, it can drive a crossover from a Kondo regime to non-Kondo regime. 

\section{Conclusion.} In summary, we have studied the CIR for two-orbital alkali-earth atoms with inter-orbital spin-exchanging interaction at finite magnetic field. We show that the CIR can strongly enhance the spin-exchanging scattering and hence dramatically increase the Kondo temperature. The signature of the Kondo effect will manifest not only in the transport properties, but also in other quantities such as spin susceptibility that can be measured by cold atom experiment \cite{Ketterle}. Our proposal is very useful to ongoing experiments on many-body physics with alkali-earth atoms.

\begin{appendix}
\section{Confinement-induced-resonance at Finite Magnetic Field}
To determine $g_{+},g_{-}$ and $g_{c}$, we can re-write the Hamiltonian in Eq.~(\ref{eqn:rel-H}) as 
\begin{align}
\hat{H}_{c}&=\left[-\frac{\hbar^{2}}{2\mu}\nabla_{{\bf r}}^{2}+\frac{\mu\omega^{2}}{2}(x^{2}+y^{2})+V_{0}\right]\left({\cal P}_{\uparrow\downarrow}+{\cal P}_{\downarrow\uparrow}\right)\nonumber\\
&+\delta{\cal P}_{\uparrow\downarrow}+V_{1}({\cal S}_{{\rm ex}}+{\cal S}_{{\rm ex}}^{\dagger}),\label{eq:ex-H}
\end{align}
where it is not necessary to consider the $|g\uparrow;e\uparrow\rangle$ and $|g\downarrow;e\downarrow\rangle$
channels at present since they  do not couple to other channels even in the presence of a
magnetic filed. Here the operators $V_{0}$ and $V_{1}$ are given by $V_{i}=2\pi\hbar^{2}a_{si}\delta({\bf r})\frac{\partial}{\partial r}(r\cdot)/\mu$
($i=1,2$) with $a_{s0}=(a_{s}^{+}+a_{s}^{-})/2$ and $a_{s1}=(a_{s}^{-}-a_{s}^{+})/2$.
In Eq.~(\ref{eq:ex-H}) we have shifted the threshold energy by a constant
$-\delta/2$ . The relative motion of incident
atoms are in the transverse ground state $\phi_{n=0,m_{z}=0}(\rho)$,
where $\rho=\sqrt{x^{2}+y^{2}}$ , $n=0,1,2,\cdots$ is the transverse
principle quantum number and $m_{z}=n,n-2,n-4,\cdots,\xi_{n}$ is
the quantum number for the angular momentum along the $z$-direction.
Here $\xi_{n}=0\ (1)$ when $n$ is even (odd). Since the system is
invariant under the rotation along the $z$-direction, $m_{z}$ is conserved and only the transverse
states with $m_{z}=0$ are involved in this problem. We further assume
that both the Zeeman energy $\delta$ and the relative kinetic energy
$\epsilon$ of the two atoms in the channel $|g\downarrow;e\uparrow\rangle$
are much smaller than the energy gap $2\hbar\omega$ between the first
transverse excited state and the transverse ground state, i.e., 
\begin{equation}
\epsilon,\delta\ll2\hbar\omega.\label{lowenergy}
\end{equation}

\begin{figure}
\includegraphics[width=3.3in]{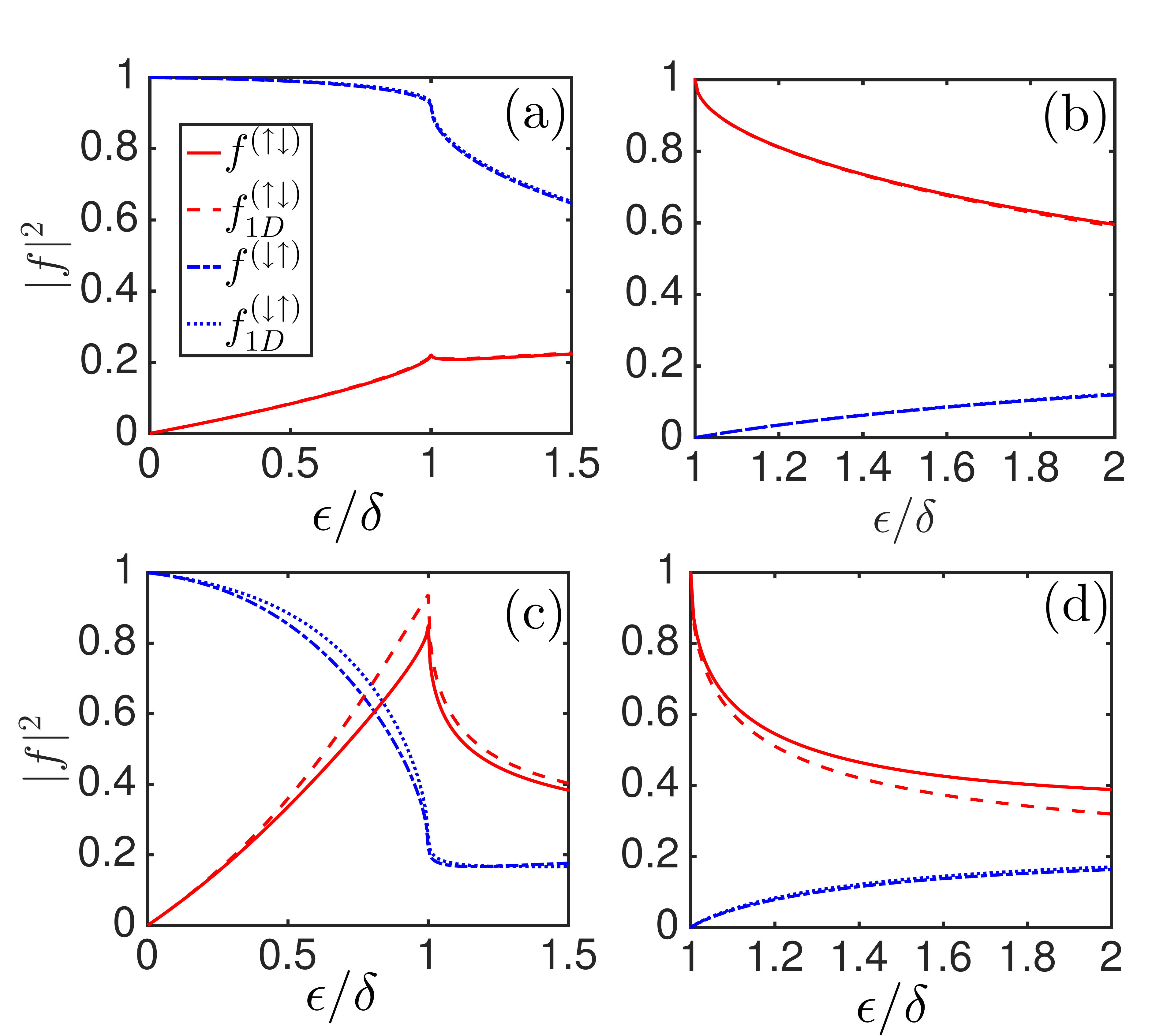} 
\caption{Scattering amplitude as function of incident energy. In (a) and (c): incident energy $\epsilon\geqslant0$ and incident atoms are in state $|g\downarrow;e\uparrow\rangle(\alpha=0,\beta=1)$. Here, $\alpha,\beta$ have been defined in Eqs.~(\ref{psiud},\ref{psidu}). In (b) and (d):  incident energy $\epsilon\geqslant\delta$ and incident atoms are in state $|g\uparrow;e\downarrow\rangle(\alpha=1,\beta=0)$. As shown in (a) and (b), $\delta\approx0.1\hbar\omega$, the scattering amplitude given by 1D model almost coincides with that given by complete calculation. While, if $\delta\approx0.4\hbar\omega$,  the difference of scattering amplitude begin to show up, as shown in (c) and (d).}
\label{fig:gvse} 
\end{figure}

In this system the two-atom scattering wave function can be written as
\begin{equation}
|\Psi({\bf r})\rangle=\Psi^{(\uparrow\downarrow)}(z,\rho)|g\uparrow;e\downarrow\rangle+\Psi^{(\downarrow\uparrow)}(z,\rho)|g\downarrow;e\uparrow\rangle,\label{psir}
\end{equation}
where the functions $\Psi^{(\uparrow\downarrow)}(z,\rho)$ and $\Psi^{(\downarrow\uparrow)}(z,\rho)$
are given by 
\begin{align}
\Psi^{(\uparrow\downarrow)}(z,\rho) & =  \left[\alpha e^{ik^{(\uparrow\downarrow)}z}+f^{(\uparrow\downarrow)}e^{ik^{(\uparrow\downarrow)}|z|}\right]\phi_{n=0,m_{z}=0}(\rho)\nonumber\\&+\sum_{n=2,4,6,8,\cdots}B_{n}^{(\uparrow\downarrow)}e^{-\kappa_{n}^{(\uparrow\downarrow)}|z|}\phi_{n,m_{z}=0}(\rho);\label{psiud}\\
\Psi^{(\downarrow\uparrow)}(z,\rho) & =  \left[\beta e^{ik^{(\downarrow\uparrow)}z}+f^{(\downarrow\uparrow)}e^{ik^{(\downarrow\uparrow)}|z|}\right]\phi_{n=0,m_{z}=0}(\rho)\nonumber\\&+\sum_{n=2,4,6,8,\cdots}B_{n}^{(\downarrow\uparrow)}e^{-\kappa_{n}^{(\downarrow\uparrow)}|z|}\phi_{n,m_{z}=0}(\rho).\label{psidu}
\end{align}
Here we consider the general case where $\epsilon$
could be either larger or smaller than $\delta$. When $\epsilon<\delta$, the incident atoms are in the state $|g\downarrow;e\uparrow\rangle$, which means $(\alpha,\beta)=(0,1)$.
When $\epsilon>\delta$, the incident atoms are either in the state $|g\uparrow;e\downarrow\rangle$ or $|g\downarrow;e\uparrow\rangle$ which means either $(\alpha,\beta)=(1,0)$ or $(\alpha,\beta)=(0,1)$ for the system. In Eqs.~(\ref{psiud},\ref{psidu})
the function $\phi_{n,m_{z}}(\rho)$ is the eigen-wave function of
the transverse Hamiltonian (i.e. the Hamiltonian of a two-dimensional
harmonic oscillator in the $x$-$y$ plane with frequency $\omega$) with
quantum numbers $(n,m_{z})$, which satisfies $\phi_{n,m_{z}=0}(\rho=0)=1/(\sqrt{\pi}a_{\perp})$
with $a_{\perp}=\sqrt{\hbar/(\mu\omega)}$ being the characteristic
length of the transverse confinement. The parameters $k^{(\uparrow\downarrow)}$,
$k^{(\downarrow\uparrow)}$, $\kappa_{n}^{(\uparrow\downarrow)}$
and $\kappa_{n}^{(\downarrow\uparrow)}$ are given by 
\begin{eqnarray}
k^{(\uparrow\downarrow)} & = & \sqrt{\frac{2\mu(\epsilon-\delta)}{\hbar^{2}}};\kappa_{n}^{(\uparrow\downarrow)}=\sqrt{\frac{2\mu\left(2n\hbar\omega+\delta-\epsilon\right)}{\hbar^{2}}};\nonumber\\
k^{(\downarrow\uparrow)} & = & \sqrt{\frac{2\mu\epsilon}{\hbar^{2}}};\ \ \ \ \ \ \ \ \kappa_{n}^{(\downarrow\uparrow)}=\sqrt{\frac{2\mu\left(2n\hbar\omega-\epsilon\right)}{\hbar^{2}}},\label{ff2}
\end{eqnarray}
while the scattering amplitudes $f^{(\uparrow\downarrow)}$ and $f^{(\downarrow\uparrow)}$
and the coefficients $B_{n}^{(\uparrow\downarrow)}$ and $B_{n}^{(\downarrow\uparrow)}$
can be obtained from the Schr$\ddot{{\rm o}}$dinger equation 
\begin{equation}
\hat{H}_{c}|\Psi({\bf r})\rangle=\left(\epsilon+\hbar\omega\right)|\Psi({\bf r})\rangle,\label{se}
\end{equation}
where the term $\hbar\omega$ in the right-hand-side of Eq.~(\ref{se})
is contributed by the zero-point energy of the transverse ground state.

\begin{figure}[tp]
\includegraphics[width=3in]{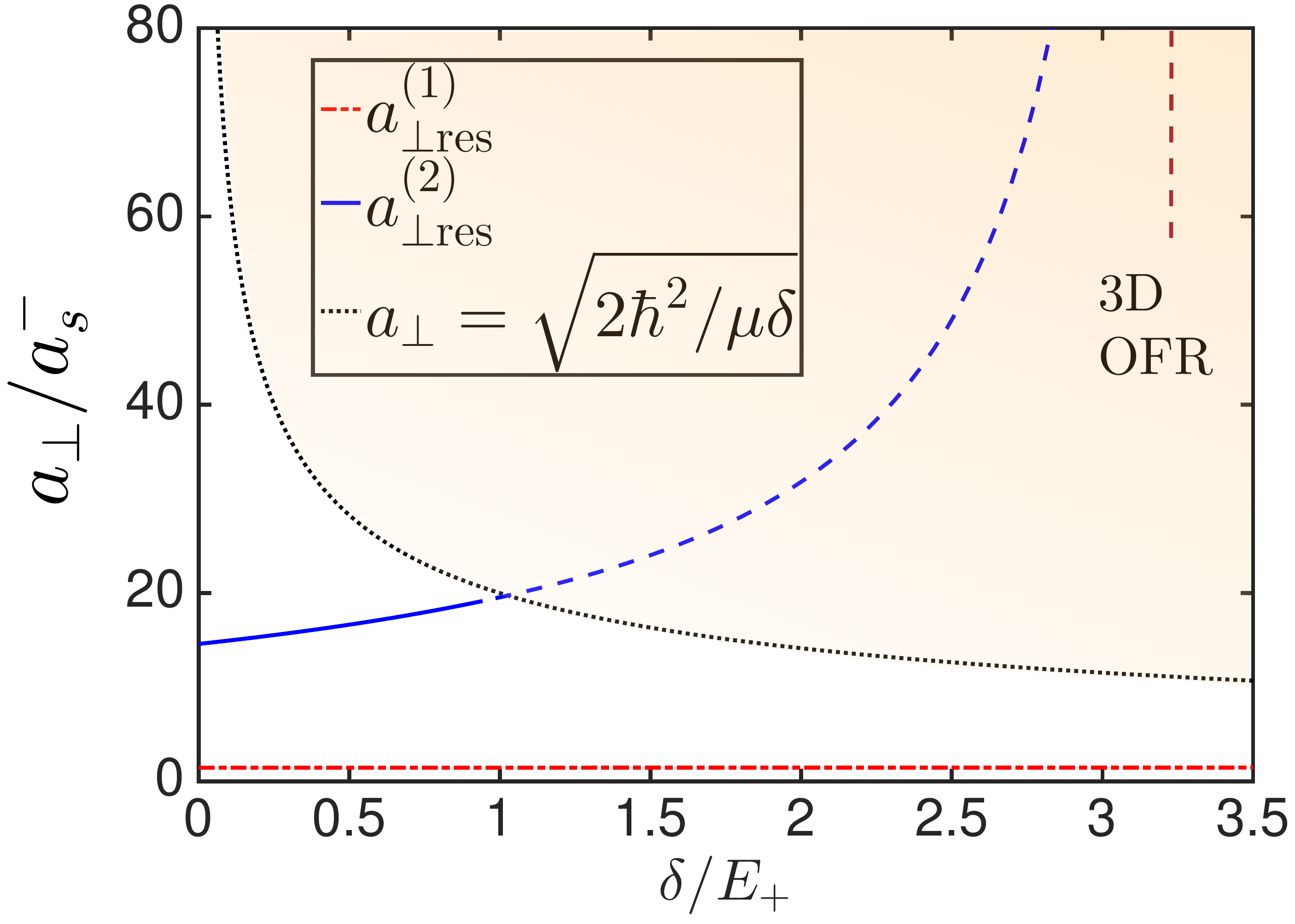} \caption{Red and blue solid line denotes the positions $a_{\perp\rm res}^{(1)}$ and $a_{\perp\rm res}^{(2)}$ of the first and second CIR, respectively. The blue dashed line denotes that if one extends the calculation to the limit of low trapping frequency, the position of the second CIR approaches the position of orbital Feshbach resonance(OFR) in 3D system \cite{OFR}, which is denoted by the red dashed line. The black dotted line is given by the condition $a_{\perp}=\sqrt{2\hbar^{2}/\mu\delta}$. Our effective 1D model is applicable only in the region with  $a_{\perp}<\sqrt{2\hbar^{2}/\mu\delta}$, i.e. the region below the dotted line.}
\label{fig:res_position} 
\end{figure}

Substituting Eqs.~(\ref{psiud},\ref{psidu}) into Eq.~(\ref{se})
and performing the operation 
\begin{eqnarray}
\frac{1}{\sqrt{2\pi}}\lim_{\varepsilon\to0}\int_{-\varepsilon}^{+\varepsilon}dz\int_{0}^{\infty}\rho d\rho\phi_{n,m_{z}=0}^{*}(\rho)\nonumber
\end{eqnarray}
on the both hands of Eq.~(\ref{se}), we can obtain the relations
\begin{align}
f^{(\uparrow\downarrow)} & =-i\frac{2\sqrt{\pi}}{k^{(\uparrow\downarrow)}a_{\perp}}\left(a_{s0}\eta^{(\uparrow\downarrow)}+a_{s1}\eta^{(\downarrow\uparrow)}\right);\label{f10}\\
B_{n}^{(\uparrow\downarrow)}&=-\frac{2\sqrt{\pi}}{\kappa_{n}^{(\uparrow\downarrow)}a_{\perp}}\left(a_{s0}\eta^{(\uparrow\downarrow)}+a_{s1}\eta^{(\downarrow\uparrow)}\right);\label{f1}\\
f^{(\downarrow\uparrow)} & =-i\frac{2\sqrt{\pi}}{k^{(\downarrow\uparrow)}a_{\perp}}\left(a_{s0}\eta^{(\downarrow\uparrow)}+a_{s1}\eta^{(\uparrow\downarrow)}\right);\label{f20}\\
B_{n}^{(\downarrow\uparrow)}&=-\frac{2\sqrt{\pi}}{\kappa_{n}^{(\downarrow\uparrow)}a_{\perp}}\left(a_{s0}\eta^{(\downarrow\uparrow)}+a_{s1}\eta^{(\uparrow\downarrow)}\right),\label{f2}
\end{align}
where 
\begin{eqnarray}
\eta^{(\uparrow\downarrow)}=\left.\frac{\partial}{\partial z}\left[z\Psi^{(\uparrow\downarrow)}(z,\rho=0)\right]\right|_{z\to0^{+}},\label{eqn:S0}\\
\eta^{(\downarrow\uparrow)}=\left.\frac{\partial}{\partial z}\left[z\Psi^{(\downarrow\uparrow)}(z,\rho=0)\right]\right|_{z\to0^{+}}.\label{eqn:S}
\end{eqnarray}
Furthermore, substituting Eqs.~(\ref{f1},\ref{f2}) into Eqs.~(\ref{psiud},\ref{psidu})
and using the fact $\phi_{n,m_{z}=0}(\rho=0)=1/(\sqrt{\pi}a_{\perp})$,
we obtain 
\begin{widetext}
\begin{eqnarray}
\Psi^{(\uparrow\downarrow)}(z,\rho=0) & = & \frac{\alpha e^{ik^{(\uparrow\downarrow)}z}}{\sqrt{\pi}a_{\perp}}-\left[a_{s0}\eta^{(\uparrow\downarrow)}+a_{s1}\eta^{(\downarrow\uparrow)}\right]\left\{ i\frac{2e^{ik^{(\uparrow\downarrow)}|z|}}{\hbar k^{(\uparrow\downarrow)}a_{\perp}^{2}}+\frac{\Lambda\left[\frac{2|z|}{a_{\perp}},-\left(\frac{k^{(\uparrow\downarrow)}}{2}a_{\perp}\right)^{2}\right]}{\hbar a_{\perp}}\right\} ;\label{psiud2}\\
\Psi^{(\downarrow\uparrow)}(z,\rho=0) & = & \frac{\beta e^{ik^{(\downarrow\uparrow)}z}}{\sqrt{\pi}a_{\perp}}-\left[a_{s0}\eta^{(\downarrow\uparrow)}+a_{s1}\eta^{(\uparrow\downarrow)}\right]\left\{ i\frac{2e^{ik^{(\downarrow\uparrow)}|z|}}{\hbar k^{(\downarrow\uparrow)}a_{\perp}^{2}}+\frac{\Lambda\left[\frac{2|z|}{a_{\perp}},-\left(\frac{k^{(\downarrow\uparrow)}}{2}a_{\perp}\right)^{2}\right]}{\hbar a_{\perp}}\right\} ,\label{psidu2}
\end{eqnarray}
\end{widetext}
where the function $\Lambda\left[\xi,\nu\right]$ is defined as $\Lambda\left[\xi,\nu\right]=\sum_{s^{\prime}=1}^{\infty}e^{-\sqrt{s^{\prime}+\nu}\xi}/\sqrt{s^{\prime}+\nu}$.
As shown in Ref.\cite{CIR1}, this function can be expanded
as 
\begin{equation}
\Lambda\left[\xi,\nu\right]=\frac{2}{\xi}+\zeta\left[\frac{1}{2},1+\nu\right]+{\cal O}(\xi),\label{lam}
\end{equation}
with $\zeta(s,a)$ being the Hurwitz-Zeta function. Substituting Eqs.~(\ref{psiud2},\ref{psidu2})
into Eq.~(\ref{eqn:S0},\ref{eqn:S}) and using Eq.~(\ref{lam}), we can obtain
the factors $\eta^{(\downarrow\uparrow)}$ and $\eta^{(\uparrow\downarrow)}$. Further using these results and Eqs.~(\ref{f10})-(\ref{f2}), we can
finally obtain the expressions of the scattering amplitudes $f^{(\uparrow\downarrow)}$
and $f^{(\downarrow\uparrow)}$:
\begin{equation}
\left(\begin{array}{c}
f^{(\uparrow\downarrow)}\\
f^{(\downarrow\uparrow)}
\end{array}\right)=\frac{-1}{I+iAP+i\sum_{s=1}^{\infty}\left(\frac{\epsilon}{2\hbar\omega}\right)^{s}A_{s}^{\prime}P}\left(\begin{array}{c}
\alpha\\
\beta
\end{array}\right),\label{fr-1}
\end{equation}
where $I,$ $A$, $P$ and $A_{s}^{\prime}$ are $\epsilon$-independent
$2\times2$ matrix with expressions 
\begin{equation}
I=\left(\begin{array}{cc}
1 & 0\\
0 & 1
\end{array}\right),\ P=\left(\begin{array}{cc}
k^{(\uparrow\downarrow)} & 0\\
0 & k^{(\downarrow\uparrow)}
\end{array}\right),\label{ik}
\end{equation}
and
\begin{equation}
A=-\frac{a_{\perp}^{2}}{2}\left[\left(\begin{array}{cc}
a_{s0} & a_{s1}\\
a_{s1} & a_{s0}
\end{array}\right)^{-1}+\left(\begin{array}{cc}
\frac{\zeta\left[\frac{1}{2},1+\frac{\delta}{2\hbar\omega}\right]}{a_{\perp}} & 0\\
0 & \frac{\zeta\left[\frac{1}{2},1\right]}{a_{\perp}}
\end{array}\right)\right].\label{am}
\end{equation}
The expressions of $A_{s}^{\prime}$ can be obtained by straightforward 
calculation we introduced above, and are not needed in the following
discussion. In Eq.~(\ref{fr-1}), $\frac{1}{[...]}$ means the inverse
matrix of $[...]$.

On the other hand, according to Eqs.~(\ref{psiud}, \ref{psidu}),
in the long-range limit $|z|\rightarrow\infty$ (i.e., $|z|\gg a_{\perp}$)
the asymptotic wave function of the two-atom relative motion reads
\begin{align}
|\Psi({\bf r})\rangle&\xrightarrow{|z|\rightarrow\infty}\Big[\left(\alpha e^{ik^{(\uparrow\downarrow)}z}+f^{(\uparrow\downarrow)}e^{ik^{(\uparrow\downarrow)}|z|}\right)|g\uparrow;e\downarrow\rangle\nonumber \\&+\left(\beta e^{ik^{(\downarrow\uparrow)}z}+f^{(\downarrow\uparrow)}e^{ik^{(\downarrow\uparrow)}|z|}\right)|g\downarrow;e\uparrow\rangle\Big]\phi_{0,0}(\rho).\label{psiasy}
\end{align}
Therefore, the long-range behavior of the two-atom relative wave function
is completely determined by Eq.~(\ref{fr-1}).

Now let us consider a pure one-dimensional system with Hamiltonian $\hat{H}_{\rm 1D}$
defined in Eq.~(\ref{eqn:eff-H}). Similar as above, in the bases
$\{|g\uparrow;e\downarrow\rangle,|g\downarrow;e\uparrow\rangle\}$,
this Hamiltonian can be re-written as 
\begin{align}
\hat{H}_{\rm 1D} & = \left(-\frac{\hbar^{2}}{2\mu}\frac{d^{2}}{dz^{2}}\right)\left({\cal P}_{\uparrow\downarrow}+{\cal P}_{\downarrow\uparrow}\right)+\delta{\cal P}_{\uparrow\downarrow}\nonumber\\&+\Bigg[\frac{g_{+}+g_{-}+2g_{c}}{2}{\cal P}_{\uparrow\downarrow}+\frac{g_{+}+g_{-}-2g_{c}}{2}{\cal P}_{\downarrow\uparrow}\nonumber\\&+\frac{g_{-}-g_{+}}{2}({\cal S}_{{\rm ex}}+{\cal S}_{{\rm ex}}^{\dagger})\Bigg]\delta(z).
\label{h1d}
\end{align}
With straightforward calculation, it is easy to find that the one-dimensional scattering
wave function in this system is given by 
\begin{align}
|\Psi_{{\rm 1D}}(z)\rangle&=\left(\alpha e^{ik^{(\uparrow\downarrow)}z}+f_{{\rm 1D}}^{(\uparrow\downarrow)}e^{ik^{(\uparrow\downarrow)}|z|}\right)|g\uparrow;e\downarrow\rangle\nonumber\\&+\left(\beta e^{ik^{(\downarrow\uparrow)}z}+f_{{\rm 1{\rm D}}}^{(\downarrow\uparrow)}e^{ik^{(\downarrow\uparrow)}|z|}\right)|g\downarrow;e\uparrow\rangle,\label{psi1d}
\end{align}
with $\alpha$, $\beta$, $k^{(\uparrow\downarrow)}$ and $k^{(\downarrow\uparrow)}$
defined the same as above, and the one-dimensional scattering amplitudes $f_{{\rm 1D}}^{(\uparrow\downarrow)}$
and $f_{{\rm 1{\rm D}}}^{(\downarrow\uparrow)}$ are given by 
\begin{equation}
\left(\begin{array}{c}
f_{{\rm 1D}}^{(\uparrow\downarrow)}\\
f_{{\rm 1D}}^{(\downarrow\uparrow)}
\end{array}\right)=-\frac{1}{I+iA_{{\rm 1D}}P}\left(\begin{array}{c}
\alpha\\
\beta
\end{array}\right),\label{fr-1-1}
\end{equation}
where $P$ is defined in Eq. (\ref{ik}) and
\begin{equation}
A_{{\rm 1D}}=-\frac{2\hbar^{2}}{\mu}\left(\begin{array}{cc}
g_{+}+g_{-}+2g_{c} & g_{-}-g_{+}\\
g_{-}-g_{+} & g_{+}+g_{-}-2g_{c}
\end{array}\right)^{-1}.\label{a1d}
\end{equation}
is the one-dimensional scattering length matrix. 

By comparing Eqs.~(\ref{fr-1},\ref{am}) with Eqs.~(\ref{fr-1-1},\ref{am}),
we find that in the systems where $\epsilon$ is much smaller than
$\hbar\omega$ so that the high-order terms $i\sum_{s=1}^{\infty}\left(\frac{\epsilon}{2\hbar\omega}\right)^{s}A_{s}^{\prime}P$
can be neglected in Eq. (\ref{fr-1}), when the one-dimensional parameters $g_{\pm}$
and $g_{c}$ satisfy 
\begin{align}
\left(\begin{array}{cc}
a_{s0} & a_{s1}\\
a_{s1} & a_{s0}
\end{array}\right)^{-1} & +  \frac{1}{a_{\perp}}\left[\begin{array}{cc}
\zeta\left(\frac{1}{2},1+\frac{\delta}{2\hbar\omega}\right) & 0\\
0 & \zeta\left(\frac{1}{2},1\right)
\end{array}\right]\nonumber\\\hspace{-4cm}&=
\frac{4\hbar}{\mu a_{\perp}^{2}}\left(\begin{array}{cc}
g_{+}+g_{-}+2g_{c} & g_{-}-g_{+}\\
g_{-}-g_{+} & g_{+}+g_{-}-2g_{c}
\end{array}\right)^{-1},\label{con}
\end{align}
we have the relation $A\approx A_{{\rm 1D}}$ which gives 
\begin{equation}
f_{{\rm 1D}}^{(\uparrow\downarrow)}\approx f_{{\rm }}^{(\uparrow\downarrow)};\ f_{{\rm 1D}}^{(\downarrow\uparrow)}\approx f_{{\rm }}^{(\downarrow\uparrow)}\label{fr}
\end{equation}
and 
\begin{equation}
|\Psi({\bf r})\rangle\xrightarrow{|z|\rightarrow\infty}\phi_{n=0,m_{z}=0}(\rho)|\Psi_{{\rm 1D}}(z)\rangle.\label{psir-1}
\end{equation}
Therefore, the Hamiltonian $\hat{H}_{\rm 1D}$ in Eq.~(\ref{eqn:eff-H})
with parameters $g_{\pm}$ and $g_{c}$ given by Eq.~(\ref{con})
is the correct  one-dimensional model for these systems and the values of $g_{+}$, $g_{-}$ and $g_\text{c}$ at finite-magnetic field are plotted in Fig. \ref{CIR_F}(b).
In Fig.~\ref{fig:gvse}, we plot $f^{(\uparrow\downarrow)}$, $f^{(\downarrow\uparrow)}$ given by complete calculation (i.e. Eq.~(\ref{fr-1})) and  $f^{(\uparrow\downarrow)}_{1D}$, $f^{(\downarrow\uparrow)}_{1D}$ from one-dimensional model (i.e. Eq.~(\ref{fr-1-1})). It is shown that as long as $\delta$ is small enough, the scattering amplitude given by one-dimensional model almost coincides with that given by complete calculation, no matter where the incident channel is located, as shown in Fig.~\ref{fig:gvse}(a,b). However when $\delta$ is comparable to $\hbar\omega$, the difference between these scattering amplitude at high-energy regime begins to show up, as shown in Fig.~\ref{fig:gvse}(c,d).

In Fig.~\ref{fig:res_position}, we show the show the characteristic length $a_{\rm res}^{(1,2)}$ for the first and second CIR as function of Zeeman energy $\delta$. The position of the first CIR is insensitive to Zeeman field, while that of the second CIR is quit sensitive to Zeeman field. It is pointed that since our calculation is based on the condition (\ref{lowenergy}), the effective one-dimensional model is applicable only in the parameter region $\delta<2\hbar\omega$ or $a_{\perp}<\sqrt{2\hbar^{2}/\mu\delta}$, i.e. the light region blow the dotted line in Fig.~\ref{fig:res_position}.

\end{appendix}

\textit{Acknowledgement.} This work is supported by Tsinghua University Initiative Scientific Research Program, NSFC Grant No. 11174176(HZ), No. 11325418 (HZ), No. 11222430 (P.Z.), No. 11434011 (PZ) and No. 11504195 (WC) and NKBRSFC under Grant No. 2011CB921500 (HZ) and No. 2012CB922104 (P.Z.).


\begin{thebibliography}{99}

\bibitem{Jun}
N. Hinkley, J. A. Sherman, N. B. Phillips, M. Schioppo, N. D. Lemke, K. Beloy, M. Pizzocaro, C. W. Oates, A. D. Ludlow, Science, {\bf 341}, 1215 (2013).
and B. J. Bloom, T. L. Nicholson, J. R. Williams, S. L. Campbell, M. Bishof, X. Zhang, W. Zhang, S. L. Bromley, and  J. Ye, Nature, {\bf 506}, 71 (2014).

\bibitem{Takahashi}
S. Taie, R. Yamazaki, S. Sugawa, Y. Takahashi, Nat. Phys. {\bf 8}, 825 (2012).

\bibitem{Fallani}
G. Pagano, M.  Mancini, G.  Cappellini, P. Lombardi, F. Sch\"afer, H. Hu, X. J. Liu, J. Catani, C. Sias, M. Inguscio and L. Fallani, Nature Phys. {\bf 10}, 198 (2014).

\bibitem{Jun1}
X. Zhang, M. Bishof, S. L. Bromley, C. V. Kraus, M. S. Safronova, P. Zoller, A. M. Rey, J. Ye, Science, {\bf 345}, 1467 (2014).

\bibitem{Munich}
F. Scazza, C. Hofrichter, M. H\"ofer, P. C. De Groot, I. Bloch, and S. F\"olling, Nature Phys. {\bf 10}, 779 (2014) and Nature Phys. {\bf 11}, 514 (2015).

\bibitem{Florence}
G. Cappellini, M. Mancini, G. Pagano, P. Lombardi, L. Livi, M. Siciliani de Cumis, P. Cancio, M. Pizzocaro, D. Calonico, F. Levi, C. Sias, J. Catani, M. Inguscio, and L. Fallani, Phys. Rev. Lett. {\bf 113}, 120402 (2014) and Phys. Rev. Lett. {\bf 114}, 239903 (2015).

\bibitem{Ray}
M. A. Cazalilla, A. M. Rey, Rep. Prog. in Phys. {\bf 77}, 124401 (2014).

\bibitem{Kondo_alkali_earth}
A. V. Gorshkov, M. Hermele, V. Gurarie, C. Xu,  P. S. Julenne, J. Ye, P. Zoller, E. Demler, M. D. Lukin and A. M. Rey, Nature Phys. {\bf 6}, 289 (2010).

\bibitem{Kondo1}
G. M. Falco, R. A. Duine, and H. T. C. Stoof, Phys. Rev. Lett. {\bf 92}, 140402 (2004).

\bibitem{Kondo2}
L. -M. Duan, Europhys. Lett. {\bf 67}, 721 (2004).

\bibitem{Kondo3}
A.  Recati, P. O. Fedichev, W. Zwerger, J. von Delft, and P. Zoller, Phys. Rev. Lett. {\bf 94}, 040404 (2005).

\bibitem{Kondo4}
B. Paredes, C. Tejedor, and J. I. Cirac, Phys. Rev. A {\bf 71}, 063608 (2005).

\bibitem{Kondo5}
P. P. Orth, I. Stanic, and K. Le Hur, Phys. Rev. A {\bf 77}, 051601 (2008).

\bibitem{Yusuke}
Y. Nishida, Phys. Rev. Lett. {\bf 111}, 135301, (2013) and arXiv: 1508.07098.

\bibitem{Demler}
J. Bauer, C. Salomon, and E.  Demler, Phys. Rev. Lett. {\bf 111}, 215304 (2013).

\bibitem{Avishai}
I. Kuzmenko, T. Kuzmenko, Y. Avishai, K. Kikoin, Phys. Rev. B {\bf 91}, 165131 (2015).

\bibitem{Rey2}
L. Isaev, A. M. Rey, arXiv: 1505.06271.

\bibitem{CIR1}
M. Olshanii, Phys. Rev. Lett. {\bf 81}, 938 (1998).

\bibitem{CIR2}
T. Bergeman, M. G. Moore, and M. Olshanii, Phys. Rev. Lett. {\bf 91}, 163201 (2003).

\bibitem{CIR_ex}
E. Haller, M. J. Mark, R. Hart, J. G. Danzl, L. Reichsollner, V. Melezhik, P. Schmelcher, and H.-C. N\"agerl, Phys. Rev. Lett. {\bf 104}, 153203 (2010).

\bibitem{Kondo}
A. C. Hewson, \textit{The Kondo Problem to Heavy Fermions}, Cambridge University Press, 1993.

\bibitem{magic_wave}
Z. W. Barber, J. E. Stalnaker, N. D. Lemke, N. Poli, C. W. Oates, T. M. Fortier, S. A. Diddams, L. Hollberg, C. W. Hoyt, A. V. Taichenachev, and V. I. Yudin, Phys. Rev. Lett. {\bf 100}, 103002 (2008).

\bibitem{Jun_PRA}
M. M. Boyd, T. Zelevinsky, A. D. Ludlow, S. Blatt, T. Zanon-Willette, S. M. Foreman, and J. Ye, Phys. Rev. A {\bf 76}, 022510 (2007).

\bibitem{OFR-exp1}
M. H\"{o}fer, L. Riegger, F. Scazza, C. Hofrichter, D.R. Fernandes, M. M. Parish, J. Levinsen, I. Bloch, S. F\"{o}lling, Phys. Rev. Lett. {\bf 115}, 265302 (2015)

\bibitem{OFR-exp2}
G. Pagano, M. Mancini, G. Cappellini, L. Livi, C. Sias, J. Catani, M. Inguscio, and L. Fallani, Phys. Rev. Lett. {\bf 115}, 265301 (2015)


\bibitem{Cui}
X. Cui, Phys. Rev. A {\bf 90}, 022705 (2014).

\bibitem{Lee}
D. H. Lee and J. Toner, Phys. Rev. Lett. {\bf 69}, 3378 (1992).

\bibitem{Nagosa}
A. Furusaki and N. Nagaosa, Phys. Rev. Lett. {\bf 72}, 892 (1994).

\bibitem{Zhai}
D. P. Zhang, W. Chen and H. Zhai,  arXiv:1510.08303.

\bibitem{Anderson}
P. W. Anderson, J. Phys. C {\bf 3}, 2436 (1970).

\bibitem{Ketterle}
C. Sanner, E. J. Su, A. Keshet, W. Huang, J. Gillen, R. Gommers, and W. Ketterle, Phys. Rev. Lett. {\bf 106}, 010402 (2011).

\bibitem{OFR}R. Zhang, Y. Cheng, H. Zhai and P. Zhang, Phys. Rev. Lett. \textbf{115}, 135301 (2015). 


\end{thebibliography}
\end{document}